\documentstyle[11pt,aaspp4]{article}

\begin{document}

\title{Optical Light Curves of the Type Ia Supernovae 1990N and 1991T}

\author{ P. Lira\altaffilmark{1}, Nicholas B. Suntzeff,
M. M. Phillips, Mario Hamuy\altaffilmark{2}\\ Electronic mail:
p.lira@roe.ac.uk, nsuntzeff@noao.edu, mphillips@noao.edu,
mhamuy@as.arizona.edu } \affil{ Cerro Tololo Inter-American
Observatory, National Optical Astronomy
Observatories,\altaffilmark{3}\\ Casilla 603, La Serena, Chile }
\and
\author{Jos\'e Maza}
\affil{
Dept. Astronom\'ia,
Universidad de Chile,
Casilla 36-D,
Santiago, Chile\\
Electronic mail: jmaza@das.uchile.cl
}
\and
\author{
R. A. Schommer, R. C. Smith\altaffilmark{4}, 
Lisa A. Wells\altaffilmark{3},\\ R. Avil\'es, J. A. Baldwin, 
J. H. Elias\altaffilmark{5}, L. Gonz\'alez, A. Layden\altaffilmark{4,6},
M. Navarrete, P. Ugarte, Alistair R. Walker, 
Gerard M. Williger\altaffilmark{7}
}
\affil{
Cerro Tololo Inter-American Observatory,
National Optical Astronomy Observatories,\altaffilmark{2}\\
Casilla 603, La Serena, Chile
}
\and
\author{
F. K. Baganoff 
}
\affil{
Center for Space Research 37-518A,
Massachusetts Institute of Technology,
77 Massachusetts Avenue,
Cambridge, MA 02139
}
\and
\author{
Arlin P. S. Crotts, R. Michael Rich, N. D. Tyson\altaffilmark{8}
}
\affil{
Department of Astronomy,
Columbia University,
Mail Code 5242,
New York, NY 10027
}
\and
\author{
A. Dey 
}
\affil{
Kitt Peak National Observatory, National
Optical Astronomy Observatories, 950 N. Cherry Ave, Tucson, Arizona 85726
}
\and
\author{
P. Guhathakurta
}
\affil{
UCO/Lick Observatory, University of California, 1156 High Street,
Santa Cruz, California 95064
}
\and
\author{
J. Hibbard
}
\affil{
Institute for Astronomy, University of Hawai'i, 2680 Woodlawn Drive, 
Honolulu, Hawai'i 96822
}
\and
\author{
Y. -C. Kim
}
\affil{
Astronomy Department, Yale University, P. O. Box 208101, New Haven,
Connecticut 06520
}
\and
\author{
Daniel M. Rehner\altaffilmark{}, E. Siciliano
}
\affil{
Space Telescope Science Institute, 3700 San Martin Drive, Baltimore,
Maryland 21218
}
\and
\author{
Joshua Roth
}
\affil{
Sky Publishing Corporation, 49 Bay State Road, Cambridge, MA 02138
}
\and
\author{
Patrick Seitzer
}
\affil{
Dept. of Astronomy, University of Michigan, 830 Dennison Bldg., 
Ann Arbor, Michigan 48109
}
\and
\author{
T. B. Williams 
}
\affil{
Physics and Astronomy Dept., Rutgers University, P. O. Box 849, 
Piscataway, New Jersey 08855
}
\altaffiltext{1}{
Current address: 
Institute for Astronomy, University of Edinburgh, Royal Observatory,
Blackford Hill, Edinburgh EH9 3HJ, UK
}
\altaffiltext{2}{
Current address: University of Arizona, Steward Observatory, Tucson,
Arizona, 85721
}
\altaffiltext{3}{
Cerro Tololo Inter-American Observatory, National
Optical Astronomy Observatories, operated by the Association of Universities
for Research in Astronomy, Inc., (AURA), under cooperative agreement with
the National Science Foundation.
}
\altaffiltext{4}{
Current address: Department of Astronomy, University of Michigan, Ann Arbor, 
Michigan 48109
}
\altaffiltext{5}{
Current address: Kitt Peak National Observatory, National
Optical Astronomy Observatories, 950 N. Cherry Ave, Tucson, Arizona 85726
}
\altaffiltext{6}{
Hubble Fellow
}
\altaffiltext{7}{
Current Address: National Optical Astronomy Observatories, Code 681,
NASA Goddard Space Flight Center, Greenbelt, Maryland 20771
USA
}
\altaffiltext{8}{
Current address: Princeton University Observatory, Peyton Hall, 
Princeton, New Jersey 08544
}
\altaffiltext{9}{
Current address: 
Harvard-Smithsonian Center for Astrophysics, 60 Garden St. Cambridge,
MA  02139
}

\begin{abstract}

We present {\em UBVRI\/} light curves for the bright Type Ia
supernovae SN 1990N in NGC 4639 and SN 1991T in NGC 4527 based on
photometry gathered in the course of the Cal\'an/Tololo supernova
program. Both objects have well-sampled light curves starting several
days before maximum light and spanning well through the exponential
tail. These data supercede the preliminary photometry published by
\cite{Lei_etal91}, and \cite{Phi_etal92}. The host galaxies for these
supernovae have (or will have) accurate distances based on the Cepheid
period-luminosity relationship. The photometric data in this paper
provide template curves for the study of general population of Type Ia
supernova and accurate photometric indices needed for the
Cepheid-supernova distance scale.

\end{abstract}



\section{Introduction}

It is now clear that Type Ia supernovae are not a homogenous class of
objects. One can see differences in spectral features at specific
epochs (\cite{Ber64}, \cite{Bra87}, \cite{Phi_etal87},
\cite{HarWhe90}, \cite{Nug_etal95}, \cite{Fil97}) and in the overall
morphology of the light curves (\cite{Phi_etal87}, \cite{Phi93},
\cite{Sun95}), which had long been suspected by earlier workers
(\cite{Bar_etal73}, \cite{Rus74}, \cite{Psk77}, \cite{Bra81},
\cite{Psk84}).  The modern data have shown us, however, that the class
of Type Ia supernovae can still be used to provide accurate {\it
relative} distances by applying correction factors to the observed
luminosity which are a simple function of the evolution of the light
curve near maximum light (\cite{Phi93}, \cite{Ham_etal95},
\cite{Rie_etal95}). The simple proof that these techniques work is the
reduction in the magnitude scatter of a Hubble diagram for distant
supernovae, where the scatter reduces from about 0.4mag to 0.14mag
(\cite{Ham_etal96b}, \cite{Rie_etal96}).

To use the very high accuracy of the zero-point of the distant
supernova Hubble diagram to measure an accurate {\it absolute}
distance scale requires the direct measurements of the distances to a
number of nearby galaxies which have had well-observed Type Ia
supernovae. One of the most accurate methods to measure absolute
distances to nearby galaxies is the use of the Cepheid
period-magnitude relationship calibrated relative to the Large
Magellanic Cloud (\cite{MadFre91}). With an independent measurement of
the distance to the LMC the observed Hubble diagram of distant
supernovae will directly yield the Hubble constant. The nearby
calibrating galaxies must have three rather trivial properties: the
galaxy must be young enough to form classical Cepheids; the galaxy
must have hosted a reasonably ``normal'' type Ia supernova; and the
supernova light curve must have been reasonably well-measured.

There are only a handful of such supernova host galaxies within the
light grasp of HST, where the limiting distance modulus for measuring
Cepheids light curves is about 32.0.  Six supernovae, SNe 1895B,
1937C, 1960F, 1972E, 1981B, and 1990N have been calibrated to date by
Saha, Sandage and collaborators (see \cite{Sah_etal96} for the most
recent paper in this series).

We have pointed out (\cite{Ham_etal96b}) that a few of the nearby
supernovae are not ideal as calibrators. The light curves for SNe
1895B and 1960F are very poor with ill-defined maxima. SN 1937C was
well observed, but the transformation from the 60 year-old films to
modern photometric bandpasses, while carefully calibrated
(\cite{PieJac95}), remains controversial in some circles
(\cite{Sch96}; rebuttal in \cite{JacPie96}). Both 1937C and 1972E
clearly had ``slower'' evolution near maximum light, which we have
shown is indicative of intrinsically brighter supernovae
(\cite{Ham_etal96a}); however, this latter point has been contested
(\cite{TamSan95}, \cite{San_etal96}). SN 1989B had very high reddening
($E(B-V) \sim 0.4$; \cite{Wel_etal94}). Only SNe 1981B and 1990N have
evidently both uncontroversial light curves and Cepheid distances.

If we relax the requirement that the Cepheids must be measured in the
{\it same} galaxy as the supernova and rely on group or cluster
associations between galaxies, a number of other calibrators become
available. Cepheid distances to NGC 3368 (M96) (\cite{Tan_etal95}) and
NGC 3351 (\cite{Gra_etal97} have been measured with HST data.
\cite{San_etal96} associate the M96 group with a larger Leo Group
(also called the Leo I cloud) which includes the compact M66 (Leo
Triplet) group. However, M66, which hosted SN 1989B, is some 8\arcdeg\
away from the Cepheid host galaxies and one must worry that the whole
Leo group is at the same distance.  HST observations to determine a
Cepheid distance to M66 are planned for HST Cycle 7 by Saha, Sandage,
and collaborators.  There have also been two well studied Type Ia
supernovae in the Fornax cluster: SN 1980N (in NGC 1316) and SN 1992A
(in NGC 1380). \cite{Sil_etal96} have measured a Cepheid distance to
the peculiar spiral NGC 1365 thought to be a member of Fornax. Once
again, the physical association of the supernova host galaxy and the
Cepheid host galaxy is a point of some controversy. Such ambiguities
lead us to prefer Type Ia supernova absolute magnitude calibrations
based on galaxies which have both primary calibrators such as Cepheids
and supernova.


SN 1990N was discovered significantly before maximum in the SBb(r)
galaxy NGC 4639 by E.~Thouvenot at the Observatory of the C\^ote
d'Azur on 22 June 1990 (\cite{Mau90}; all dates referenced as UT).
\cite{Pol90} measured an astrometric position for this supernova of
(RA,dec,equinox)= (15$^h$ 18$^m$ 52\fs92, -7\arcdeg 11\arcmin
43\farcs2, 1950). \cite{KirLei90} classified it as a Type Ia supernova
on the basis of a spectrum obtained on the 26 June
1990. \cite{Lei_etal91} presented preliminary light curves based on
CTIO CCD data (which is reanalyzed in the present paper).  Spectral
modeling has been published by \cite{Jef_etal92}, \cite{Shi_etal92},
\cite{Yam_etal92}, and \cite{Fis_etal97}. Recently, \cite{San_etal96}
and \cite{Sah_etal96} have obtained the distance to the parent galaxy
by measuring the periods and magnitudes of 20 Cepheid variable stars
with the Hubble Space Telescope. This object is therefore a key
template objects in establishing the value of $H_0$ based on the
Cepheid-supernova distance scale.

SN 1991T has been one of the most extensively studied Type Ia
supernovae. It was discovered well before maximum in the Sb(s)II
galaxy NGC 4527 by S. Knight and independent observers
(\cite{Waa_etal91}) on 13 April 1991.  An astrometric position for
this supernova by R. H. McNaught is given in the previous reference as
(RA,dec,equinox)= (12$^h$ 31$^m$ 36\fs91, +2\arcdeg 56\arcmin
28\farcs3, 2000).

Early optical spectral observations reported by \cite{LaFGol91},
\cite{Kir91}, and \cite{PhiHam91} showed that SN 1991T was a peculiar
Type Ia event which motivated a number of theoretical studies to model
the spectral evolution (\cite{Jef_etal92}, \cite{Rui_etal92},
\cite{Spy_etal92}, \cite{Yam_etal92}, \cite{Maz_etal95},
\cite{Mei_etal96}). Optical photometry has been published by
\cite{Phi_etal92}, \cite{For_etal93}, and \cite{Sch_etal94} which
showed that SN 1991T had a very slow rate of evolution through
maximum.  Due to the excellent temporal coverage of the light curve,
this supernova has been used as template example of a slow supernova
(\cite{Ham_etal96c}, \cite{Rie_etal96}). It is expected that the Saha
and Sandage group will obtain a Cepheid distance to NGC 4527 using
data taken with HST in Cycle 7.

When accurate modern light curves for several nearby supernovae became
available some years ago, subtle differences between Type Ia supernova
light curves became apparent.  CCD photometry showed that there is a
real spread in the peak luminosity and that some of the objects evolve
through maximum light more slowly than others.  \cite{Phi93} presented
evidence that the rate of the decline after maximum is correlated with
the luminosity at maximum and that more luminous objects have a slower
decline rates.  \cite{Ham_etal95}, \cite{Rie_etal95},
\cite{Ham_etal96b}, and \cite{Rie_etal96} have found that the scatter
in the Hubble diagram of Type Ia supernovae decreases significantly
when corrections for the peak luminosity -- decline rate relation are
introduced.  \cite{TamSan95} argue that when samples are restricted to
``normal'' objects (by eliminating events like SN 1991T) there is no
need to correct for a peak luminosity -- decline rate effect.
However, the Hamuy studies find that by ignoring this effect, the
estimate of the Hubble constant can be biased too low by up to 15\%.

The intent of this paper is to present accurate light curves of the
two nearby supernovae SNe 1990N and 1991T which are important
calibrators in the distance scale.  We have already used these light
curves in our work on the Hubble constant (\cite{Ham_etal96b}. In
Section 2 of this paper we present the observations and reduction of
the optical photometric data obtained at CTIO. The light curves as
well as color curves for both supernovae are shown in Section 3. A
final discussion is found in Section 4.

\section{Observations}

The optical observations of SNe 1990N and 1991T were obtained using
the 0.9m and the Blanco 4m telescopes at CTIO.  SN 1990N was
observed from June 1990 to March 1992 and SN 1991T was observed from
April 1991 to June 1992. The observations were made using Texas
Instrument and Tektronix CCDs (except for the night of the 19 March
1990 when a Thomson detector was used) and facility $UBV(RI)_{KC}$
filters in the Johnson--Kron--Cousin photometric system (\cite{Joh63},
\cite{Kro53}, \cite{Cou76}. The observation logs are given in Table
\ref{t1} and \ref{t2}. The detector name, listed in the final
column of these two tables, combines the manufacturer name and a
running number assigned by the CTIO CCD lab. We have assumed that each
different CCD listed in this table (along with the filter set) has a
unique set of color terms that must be derived from observations.
 
We made observations of the supernovae under varied photometric
conditions, including very non-photometric weather with cloud
extinction up to a few magnitudes.  It is well established
(\cite{Ser70}, \cite{Wal_etal70}, \cite{Ols83}) that clouds are quite
grey, which allows us to use local standards (in the same CCD frame as
the supernova) and averaged color terms for a specific CCD measured on
photometric nights.

For accurate photometry on non-photometric nights, it is necessary to
define a precise local photometric sequence of stars.  We measured
photometric sequences on 13 photometric nights in the CCD field around
NGC 4639 and NGC 4527 referenced to the Landolt and Graham standards
stars (\cite{Lan72}, \cite{Gra82}, \cite{Lan92}).  Extinction
coefficients, color terms and zero points for the transformations to
the standard $UBV(RI)_{KC}$ system were derived for each night
following the method described by \cite{Har_etal81}.  Typical values
of the extinction coefficients were $k_{U}=0.50$, $k_{B}=0.32$,
$k_{V}=0.20$, $k_{R}=0.14$ and $k_{I}=0.08$ in units of mag
(airmass)$^{-1}$.
\footnote{These extinction values are higher than normal due to the
effects of the Mt.~Pinatubo eruption which occurred on JD 2448422. See
\cite{GroGoc92}.}  We measured $UBV(RI)_{KC}$ sequences for a total of
15 stars for SN 1990N and 9 stars for SN 1991T using digital aperture
photometry with an aperture diameter of 14\arcsec.  The photometric
sequences are identified in Figures \ref{f1} and \ref{f2}, and the
photometry is given in Tables \ref{t3} and \ref{t4}. In these tables
we list the number of observing nights (n) and the total number of
observations (m).

Star 2 in our local sequence around SN 1991T is also a sequence star
(number 2) listed by \cite{For_etal93}.  The magnitude differences for
this star in the sense of this study {\it minus} \cite{For_etal93} are
$\Delta(VRI) = (0.00,-0.02,-0.03)$. For the three SN 1991T local
standards in common with \cite{Sch_etal94}, we find the mean
differences in the sense of this study {\it minus} that of
\cite{Sch_etal94} are $\Delta(BVRI) = (-0.03\pm0.01, -0.04\pm0.01,
-0.02\pm0.02, +0.12\pm0.08)$ where the errors quoted are the mean
errors. These mean differences are consistent with the photometric
errors quoted in \cite{Sch_etal94}, which dominate the comparison of
the two magnitude systems. Due to the larger number of measurements on
independent photometric nights, we are confident that sequences given
in Tables \ref{t3} and \ref{t4} are the most accurate available.

To determine the supernova magnitudes we subtracted late-time images
of the parent galaxies at the location of the supernovae using the
technique described by \cite{Ham_etal94}. For these subtractions deep
``master images'' of NGC 4639 and NGC 4527 were obtained at the
beginning of 1994, which corresponds to 1300 and 1010 days after
maximum light for SNe 1990N and 1991T. A simple extrapolation of the
late-time decline rates (see the $\gamma$ parameter in Table~\ref{t8})
to these dates yields $B$ magnitudes of $\sim33$ (SN 1990N) and
$\sim28$ (SN 1991T).  However, there are two factors which could make
the late-time magnitudes in the master images significantly brighter
than this extrapolation

The first factor is the presence of minor radioactive nuclides and the
efficiency of positron energy deposition from the radioactive decays.
A Type Ia explosion is predicted to synthesize about 0.5M\sun\ of
$^{56}$Ni, and smaller amounts of $^{56}$Co, $^{57}$Co, $^{44}$Ti, and
$^{22}$Na (see model W7 in \cite{Nom_etal84}). \cite{Woo_etal89}
provide energy deposition rates for these nuclides. A complication
arises in predicting the late-time light curve after day 500 in how to
handle the energy deposition from the positron production
(\cite{Arn79}). The positrons can add energy into the supernova nebula
both from their kinetic energies and annihilations. The efficiency of
this process is poorly understood in the low-density environment of
the supernova nebula at late-time. If we make the rather extreme
assumption of complete kinetic energy deposition and positron
annihilation into gamma rays, we can use the deposition rates given by
\cite{Woo_etal89} and the model W7 abundances by \cite{Nom_etal84} to
predict upper limits to the supernova luminosity at late time.  The
effect of full energy deposition from positrons and the existence of
long-lived radioactive nuclides such as $^{44}$Ti, and $^{22}$Na tends
to flatten out the light curve past day 1000. Using a column depth of
400 g cm$^{-2}$ at t=$10^6$ s as suggested by \cite{Woo_etal89} for a
Type Ia supernova, we predict $V$ magnitudes of 29.6 and 25.4 for SNe
1990N and 1991T at the epoch of the master image. For no positron
energy deposition, the predicted magnitudes are about 3 magnitudes
fainter. In either case, the predicted magnitudes in the master images
are so faint as to have no effect on the measured photometry. However,
if there is significant overproduction of $^{57}$Co or $^{44}$Ti
relative to model W7, the late-time magnitudes could be much brighter
and affect the magnitudes measured by image subtraction.

The second factor that could affect the magnitudes measured from the
subtracted images is the presence of a light echo.  In the late-time
images of SN 1991T, the location of supernova has been found to be
contaminated by a faint echo of SN 1991T at maximum light
(\cite{Sch_etal94}). We will return to this minor complication in
Section 3.

We measured differential photometry of the supernovae on each CCD
frame using aperture photometry when the supernova was bright, or
using the point spread function (psf) fitting program DAOPHOT
(\cite{Ste87}) when the supernova was faint. Averaged color terms
chosen to match the CCD/filter setup of instruments for each observing
night were adopted from a database of coefficients at CTIO.  For SN
1991T near maximum, our CCD exposures were very short and the local
standards were poorly exposed. In this case we used the sharp core of
NGC 4527 as a $BV$ ``standard'' for the nights of 26, 28, 29, and 30
April, and 1 May 1991. An aperture radius of 2.7\arcsec\ was chosen to
maximize the signal to noise ratio for the photometry of the core. The
core photometry is listed in Table \ref{t4}.

Besides the error given by the Poisson statistics of the number of
counts in the supernova aperture or psf, $\sigma_{phot}$, there are
other error sources such as those due to the transformation of the
instrumental magnitudes into the standard system and CCD flat
fielding. To get a sense of the magnitude of these errors we selected
many frames with a sufficient number of bright stars (with negligible
$\sigma_{phot}$ error). For each frame we calculated the standard
deviation of the difference between a given measurement and the
standard magnitudes listed in Tables \ref{t3} and \ref{t4}.  This
standard deviation ($\sigma_{rms}$) is an empirical estimate of the
average error in a {\it single} observation of a stellar object in any
CCD frame when referenced to the local photometric sequence.  The
measured standard deviations $\sigma_{rms}$ were (0.026, 0.017,
0.017, 0.015, 0.017) magnitudes in $UBVRI$. The value of
$\sigma_{rms}$ derived for both supernovae in each filter agreed
within 0.002 magnitudes).

The final error in the individual magnitudes of the supernovae was
calculated as the quadratic sum of the empirical error in a single
observation $\sigma_{rms}$ and the photon statistical error,
$\sigma_{phot}$. The error $\sigma_{rms}$ was the dominant component
of the errors in the early part of the photometry, while
$\sigma_{phot}$ became more important when the supernova dimmed.

\section{Results}

\subsection{Light Curves}

We present $UBV(RI)_{KC}$ photometry for SNe 1990N and 1991T in
Tables \ref{t5} and \ref{t6}, and plot the data in Figures \ref{f3}
and \ref{f4}.  To find the time and magnitude of maximum light for
both supernovae we fit the data around the peak with a third-order
polynomial. For SN 1990N the first observation was acquired 11 days
before $B_{max}$, and the last observation was made 607 days after
$B_{max}$ in just the $V$ band. The $B$ maximum was reached on JD
2448082.7 $\pm$ 0.5 with $B_{max} = 12.76 \pm 0.03$ and a $B-V$ color
of 0.03 magnitudes.  For SN 1991T the data begin 12 days before
$B_{max}$ and end 401 days after maximum. We derive $B_{max} = 11.70
\pm 0.02$ at JD 2448375.7 $\pm$ 0.5 with a $B-V$ color of 0.17. In
Table \ref{t7} we summarize the maxima of the light curves in the
different bands for both supernovae. We find that the $B$ maximum
occurs before the $V$ maximum; in particular, the time difference
between the $B$ and $V$ maxima is $1.5 \pm 0.7$ days for SN 1990N and
$2.6 \pm 0.7$ days for SN 1991T, in agreement with the result of
\cite{Lei88} who found a difference of $2.5 \pm 0.5$ days.

Comparisons of our $BV$ photometry with previous data published for SNe
1990N and 1991T are shown in Figures \ref{f5} and \ref{f6}. The
agreement between the photometry presented in this paper and the
results published by \cite{Lei_etal91} for SN 1990N and
\cite{Phi_etal92} for SN 1991T are not surprising since they are based
on a subsample of the same optical data analyzed here.  However, a
small systematic difference between the different data sets is
clear. Our photometry near maximum is generally dimmer, although the
discrepancy is less than 0.1mag for SN 1990N and even less for SN
1991T. The preliminary photometric results of these earlier papers
which were based on only a single night for the photometric
calibration should be ignored in preference to the photometric data
given in Tables \ref{t5} and \ref{t6}.

For SN 1991T there is independent photometry published by
\cite{For_etal93} which we plot in Figure \ref{f6}. If we interpolate
our data to the dates of the \cite{For_etal93} data using a spline
fit, we find the following differences in the sense of this work {\it
minus} Ford {\it et al.}: $V, 0.08 \pm 0.01$ ; $R, 0.04 \pm 0.02$ ;
and $I, 0.02 \pm 0.02 $. The quoted errors are the errors in the mean
based on the interpolation to the 12 dates in the Ford {\it et al.}
study. A similar systematic offset in the $V$ magnitude was noted by
Ford {\it et al.}  with respect to the \cite{Phi_etal92} reductions of
the 1991T data. These mean differences between careful photometric
studies indicate the level in systematic errors that can be
encountered even in bright supernova photometry.

It is now well established that there is not a unique light curve for
all type Ia supernova. As was suggested by \cite{Psk77} and
\cite{Bra81}, supernovae can be discriminated by the rate of decline
after maximum. Pskovskii(1977, 1984) defined the parameter $\beta$ as
the characteristic decline rate during the fast-decline phase of the
$B$ supernova light curve, and the parameter $\gamma$ as the rate
during the slow-decline phase (see \cite{Phi_etal87} for an
unambiguous description of these parameters).  \cite{Phi93} introduced
the parameter $\Delta m_{15}$, defined as the decline in magnitude
during the 15 days after $B$ maximum. The evidence from nearby
supernovae (\cite{Phi93}, \cite{Ham_etal96c}) and the scatter in the
observed Hubble diagram
(\cite{Maz_etal94}, \cite{Ham_etal95}, \cite{Ham_etal96b}) clearly show
that the brighter supernova decline more slowly (small $\Delta
m_{15}$).

In Table \ref{t8} we list the evolutionary parameters of the $B$ light
curves for our two supernovae. We also list the values for the
\cite{Lei88} template $B$ curve.  This multicolor template, which was
formed from a large number of supernova light curves, provides a
useful fiducial light curve and can be considered a ``typical'' light
curve which can be compared to other observations.  We calculated the
Pskovskii $\beta$ and $\gamma$ parameters for the $B$ curves of
SNe 1990N and 1991T using a linear least-squares fit with the data
weighted using the photometric errors as quoted in Tables \ref{t5} and
\ref{t6}. The range of days used in the fitting for each supernova are
indicated in Table \ref{t8}.

Table \ref{t8} shows that in both the $\beta(B)$ and $\Delta m_{15}$
are much smaller for SN 1991T than the values for the Leibundgut
template. \cite{Phi93} has shown that this ``slow'' supernova was
intrinsically very bright. In fact, SN 1991T is one of the slowest
supernovae ever found and has been used as a representative template
for slow events (\cite{Ham_etal95}, \cite{Ham_etal96c}). SN 1990N, on
the other hand, is quite similar to the Leibundgut template, and is
therefore similar to the typical Type Ia event.

In Figures \ref{f7} and \ref{f8} we plot the first 120 days of $BV$
photometry for the two supernovae along with the Leibundgut templates.
The $BV$ templates have been shifted to match the epoch of $B$ maxima
and the peak magnitudes given in Table \ref{t7} (with the appropriate
time delay between $B$ and $V$ maximum given above). The results given
in the preceeding paragraph can be now clearly seen in these figures.
SN 1990N follows the template closely while SN 1991T declines from
maximum light more slowly. SN 1991T also begins its exponential
decline significantly earlier and remains at higher relative
brightness when compared to the template. Visually, the ``knee'' in
the light curve around 30 days past maximum occurs earlier in this
supernova. SN 1991T also rises to maximum light more slowly than SN
1990N.

In Figure \ref{f12} we plot the late-time photometry of SN 1991T from
this study and \cite{Sch_etal94} along with the predicted trend of the
late-time evolution based on the $\gamma(B)$ and $\gamma(V)$ fits to
our data. The fact that the light curve past day 400 levels off has
been shown by \cite{Sch_etal94} to be due to a light echo with $BV
\sim (21.3,21.4)$. Recall that in our work we subtracted a late-time
image of the region taken around JD2449380 around the supernova to
remove the galaxian contribution to the background under the psf. By
doing this however, we also automatically correct for the echo
contamination. This assumption is valid provided that the echo
magnitude did not change during the period between the supernova
observations and the late-time image, and that the light curve of
supernova did not level off for other reasons, such as the
overproduction of $^{44}$Ti or $^{22}$Na. Under these assumptions, the
photometry of SN 1991T in Table \ref{t6} and Figure \ref{f12} should
be free of any echo contamination. Indeed, the small differences
between our last points and those of Schmidt {\it et al.} at JD
2448750 shown in Figure \ref{f12} are consistent with the echo
magnitude cited above.

\subsection{Color Curves}

The $B-V, B-R, B-I, V-R, V-I, R-I$ color curves for SNe 1990N and 
1991T through day 100 are shown in Figures \ref{f9} and \ref{f10}
respectively. The temporal axis was shifted so that the $B$ maximum
corresponds to $t=0$ for both supernovae. The redder color of SN 1991T
with respect to SN 1990N is evident. The presence of redshifted Na
absorption lines in the spectrum of SN 1991T and the location of the
supernova in one of the arms of NGC 4527 suggest that this object was
obscured by dust in its parent galaxy. Strong Ca and Na interstellar
absorption lines at the radial velocity of NGC 4527 were observed by
\cite{WheSmi91} and \cite{MeyRot91}.  \cite{Rui_etal92} estimated an
excess $E(B-V) \sim 0.3$ assuming a relationship between the
equivalent width of the line Na I D and $E(B-V)$. On the other hand,
\cite{Phi_etal92} found an excess of 0.13 magnitudes assuming an
intrinsic color $(B-V)$ of zero during maximum. The foreground
$E(B-V)$ reddening is $0.00 \pm 0.015$ according to \cite{BurHei94}.

SN 1990N did not show absorption lines in a low-dispersion spectrum
(\cite{Lei_etal91} and its location in the outskirts of NGC 4639
suggests that this object is less reddened than SN 1991T.
\cite{Sah_etal96} estimate the mean extinction of the Cepheids as
$E(V-I)=0.04\pm0.06$ based on the difference in distance moduli from
Cepheid $VI$ P-L relations.  The foreground $E(B-V)$ reddening is
$0.012 \pm 0.015$ according to \cite{BurHei94}. It is unfortunate that
no high-dispersion spectrum of this bright object was made. The
$(B-V)$ color of 0.03 is consistent with an intrinsic color at maximum
of $\sim -0.1 - 0.1$ magnitudes for other Type Ia supernovae with low
reddenings (\cite{Ham_etal91}, \cite{SanTam93}, \cite{Ham_etal95}) and
suggests $E(B-V) \lesssim 0.15$.  \cite{Lir96} has shown that the
color evolution in $BV$ from days 32-92 during the nebular phase is
extremely uniform among Type Ia supernovae. In a future paper we will
use this fact to calibrate the intrinsic colors of SNe Ia at maximum
which, in turn, should allow more precise estimates to be made of the
host galaxy reddening.

Differences in the color evolution of the two supernovae are better
appreciated in the color-color plot shown in Figure \ref{f11}.  Time
along the light curve is indicated by labeling the points at
approximately -10, 0, 10 and 20 days from the maximum. The figure also
shows the reddening vector for a galactic extinction law
(\cite{SavMat79}).  For $t>10$ days, the curves are parallel to the
reddening vector.  However, the data from $t=-10$ to $t=10$ show that
the color curve of SN 1991T cannot be matched to that of 1990N by a
simple dereddening vector.

\section{Concluding Remarks}

SNe 1990N and 1991T are important supernovae. They were close enough
that distances to the host galaxies can (or will) be measured by
direct techniques with the HST. The light curves are especially well
determined over the full evolution, and in particular, the evolution
before maximum is well covered. The light curves of these supernovae
have become standard templates used in the study of more distant
supernovae.

The photometric data presented in this paper show that SN 1990N was a
typical Type Ia event in that the light curves are well fitted by the
template curve determined by \cite{Lei88}.  It also falls in the
middle of the range of $\Delta m_{15}$ types defined by \cite{Phi93}.
The spectral evolution of SN 1990N has also been classified as similar
to other prototypical Type Ia supernovae, although the early first
observations made it to be claimed as a peculiar object
(\cite{Lei_etal91}).

The preliminary reductions of the SN 1990N data in \cite{Lei_etal91}
have been used by \cite{San_etal96} to estimate a peak absolute
magnitude for this supernova. The peak magnitudes cited by Sandage et
al. of $(B,V)=(12.70,12.61)$ are $\sim 0.07$ mag brighter than the
more precise results given in Table \ref{t7}. Such a small magnitude
difference will have little effect on the measurement of the Hubble
constant since $\delta{H_0}/H_0 \approx
0.46\delta{m}$. \cite{Ham_etal96b}, \cite{San_etal96} and
\cite{Rie_etal96} have used the light curve of this supernova as one
of the fundamental calibrators of the absolute magnitudes of Type Ia
supernovae.  These absolute magnitudes coupled with the observed
Hubble diagram from the Cal\'an/Tololo survey (\cite{Ham_etal96b})
have yielded $H_0 \sim 65$ km s$^{-1}$ Mpc$^{-1}$.

Because of its peculiar nature, SN 1991T has been studied intensively.
The peculiarities of this supernova include pre-maximum spectra
dominated by iron-group features, a very small $\Delta m_{15}$ value,
and a visual luminosity larger than other typical Type Ia supernovae,
although the derived absolute magnitudes depend strongly on the
different extinction assumed for the supernova and the distance to the
host galaxy NGC 4527 (\cite{Fil_etal92}, \cite{Rui_etal92}, 
\cite{Phi_etal92}).  The results of this paper confirm the slow
evolutionary rate near maximum and also show that the color curve is
significantly different from more normal Type Ia supernovae.

This is not to say that SN 1991T is unique, as new events of this
``slow-class'' have been found, such as SNe 1991ag
(\cite{Ham_etal95}), 1992bc (\cite{Maz_etal94}), 1995ac
(\cite{Gar_etal96}) or 1997br (\cite{Qia_etal97b}).  The evidence
suggests that the decline rate of these supernovae is just the slow
end of the peak luminosity -- decline rate relation for Type Ia
supernovae and that this correlation could be also extended to a
spectroscopic sequence (\cite{Nug_etal95}).  \cite{Ham_etal96c} and
\cite{Gar_etal96} however, have pointed out that the intrinsic
luminosity, spectral features, and colors at maximum light are not a
simple function of the light curve shape (as measured by $\Delta
m_{15}$) for this bright class of supernovae.  For instance, among
supernovae with similar small values of $\Delta m_{15}$, SNe 1991T,
1995ac, and 1997br had very weak Si II 6355\AA\ at maximum light
(\cite{FilLeo95}, \cite{Qia_etal97a}) while 1992bc had the typical
deep spectral features at maximum light common to most Type Ia events
(\cite{Maz_etal94}). Conversely, SN 1995bd had a spectrum similar to
1991T at maximum light but its light curve was well fit by the
``faster'' Leibundgut template (\cite{Gar_etal96}).  It is clearly
important to obtain more examples of this class of bright Type Ia
supernovae to sort out this issue.

\acknowledgments JM and MH acknowledge support by Ca\'tedra
Presidencial de Ciencias 1996-1997. We would like to thank the Space
Telescope Science Institute for access to the Digitized Sky Survey. We
thank Peter Garnavich, Eric Olsen, Brian Schmidt, and Gordon Walker
for helpful correspondence. This research has made extensive use of
the Canadian Astronomy Data Center (Dominion Astrophysical
Observatory, Herzberg Institute of Astrophysics), and the NASA
Astrophysics Data System Abstract Service.  We would also like to
thank Brian Marsden and Daniel Green at the IAU Central Bureau for
Astronomical Telegrams for their valuable notification service which
allows observers to start observing supernovae within 24 hours of
discovery.



\clearpage

{}

\newpage

\figcaption[plira.fig1.ps]{SN 1990N in NGC 4639. The local standards
listed in Table \ref{t3} are marked. This $R$ image was taken at the
CTIO 0.9m telescope on 13 March 1994. The field is 6.8\arcmin\ on a
side. Star 14 does not appear on this chart. It is located at
(12:42:47.0,+13:17:52, J2000) on the Digitized Sky Survey. \label{f1}}

\figcaption[plira.fig2.ps]{SN 1991T in NGC 4527. The local standards
listed in Table \ref{t4} are marked. This $R$ image was taken at the
CTIO 0.9m telescope on 14 March 1994. The field is 6.8\arcmin\ on a
side. \label{f2}}

\figcaption[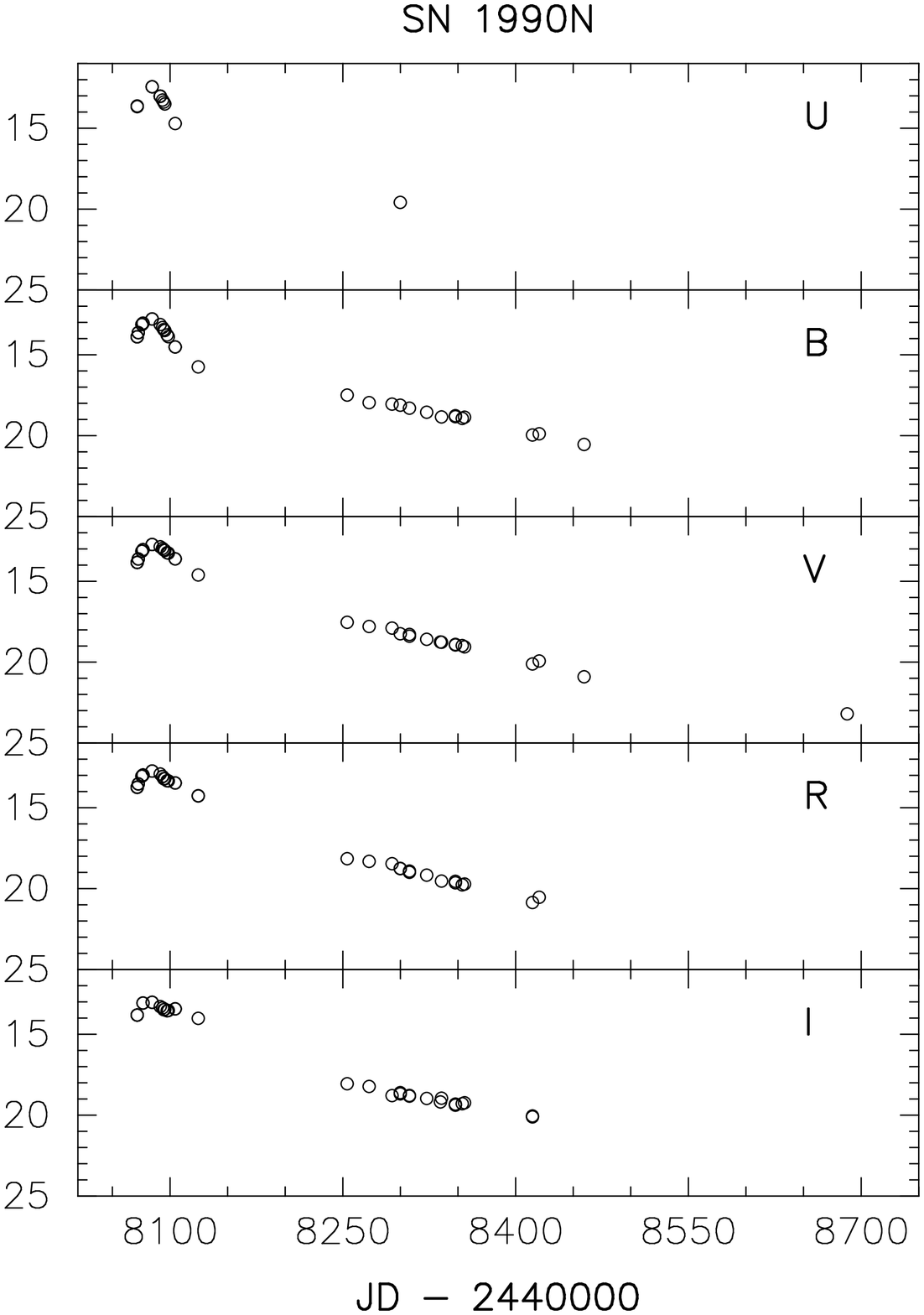]{$UBVRI$ light curves of SN 1990N.  \label{f3}}
  
\figcaption[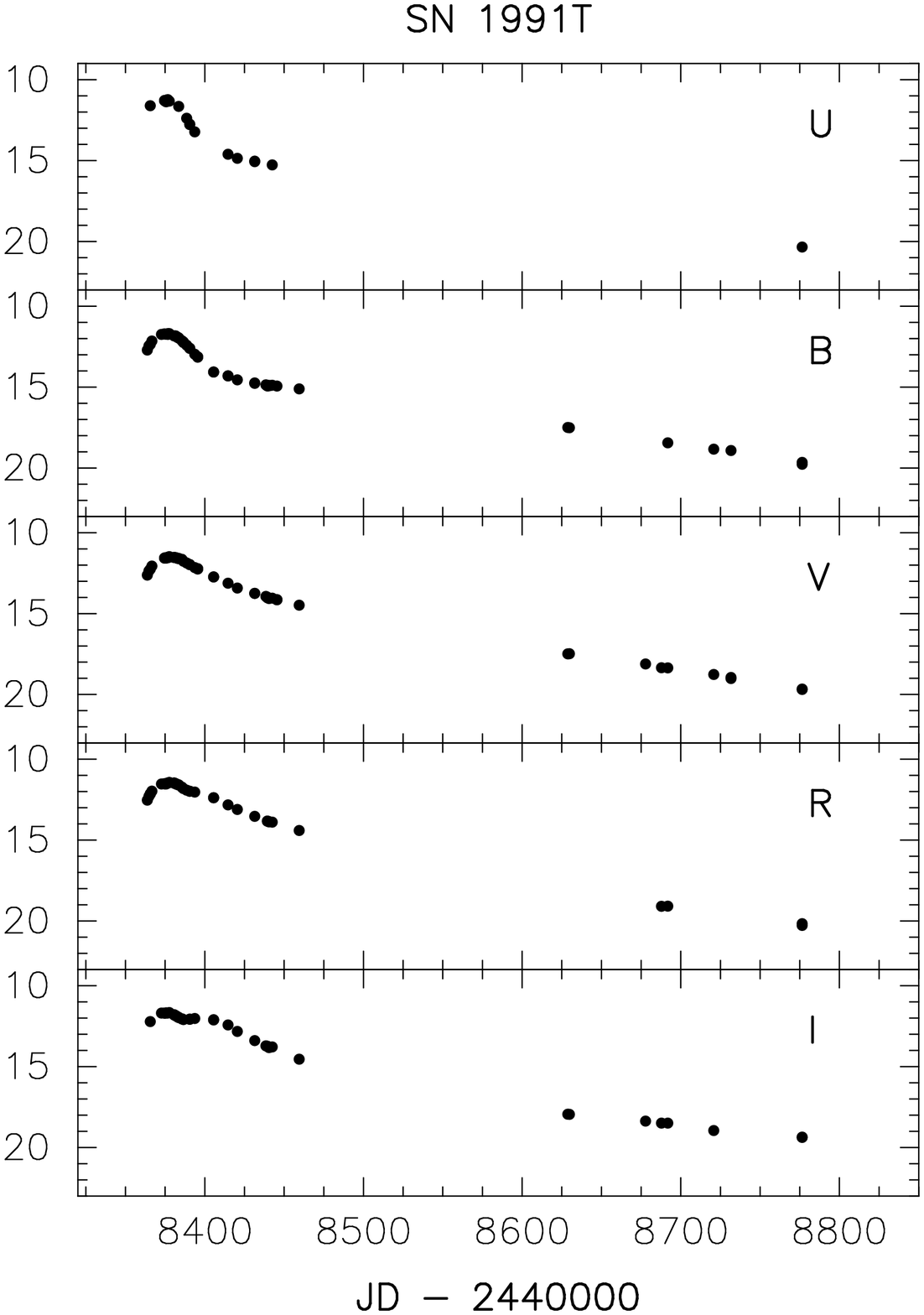]{$UBVRI$ light curves of SN 1991T. \label{f4}}

\figcaption[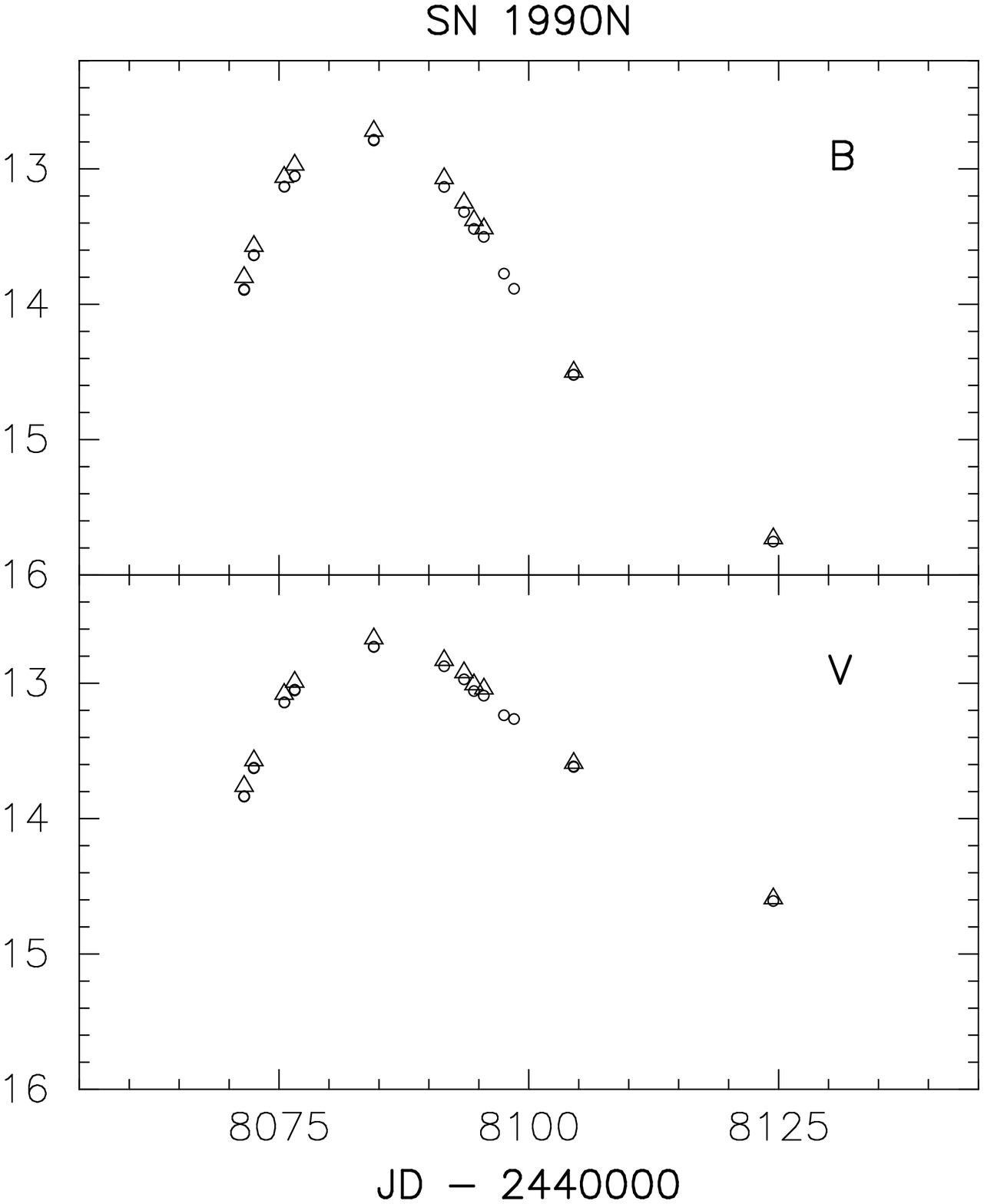]{Comparison of the $B$ and $V$ photometry
of SN 1990N given in this paper (circles) with the preliminary
photometric reductions of the CTIO data by
\protect{\cite{Lei_etal91}}(triangles). \label{f5}}

\figcaption[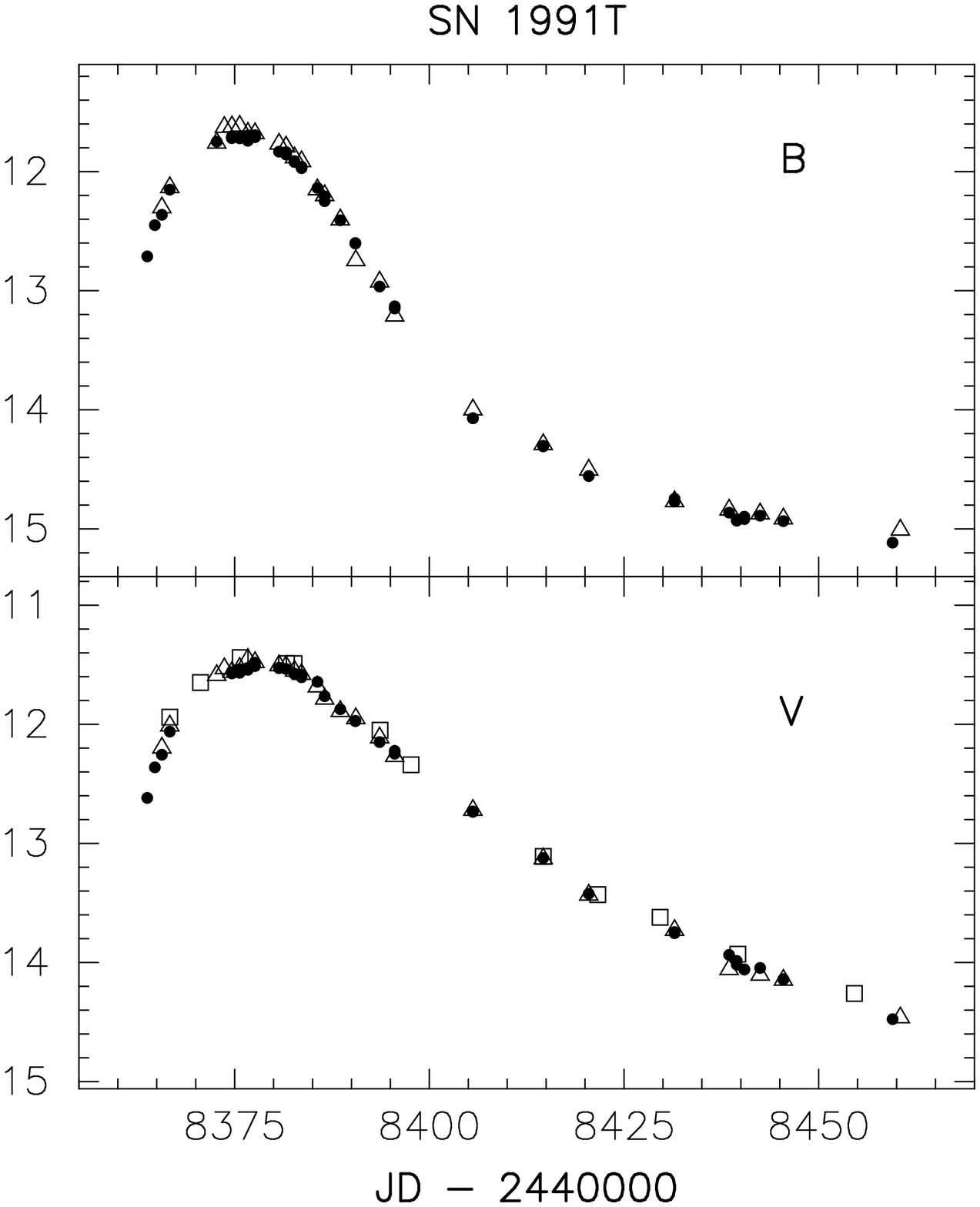]{Comparison of our $B$ and $V$ light
curves of SN 1991T (filled circles) with the preliminary photometric
reductions of the CTIO data published by \protect{\cite{Phi_etal92}}
(triangle). The $V$ magnitudes from the Van Vleck Observatory study
\protect{\cite{For_etal93}} are plotted as squares. \label{f6}}

\figcaption[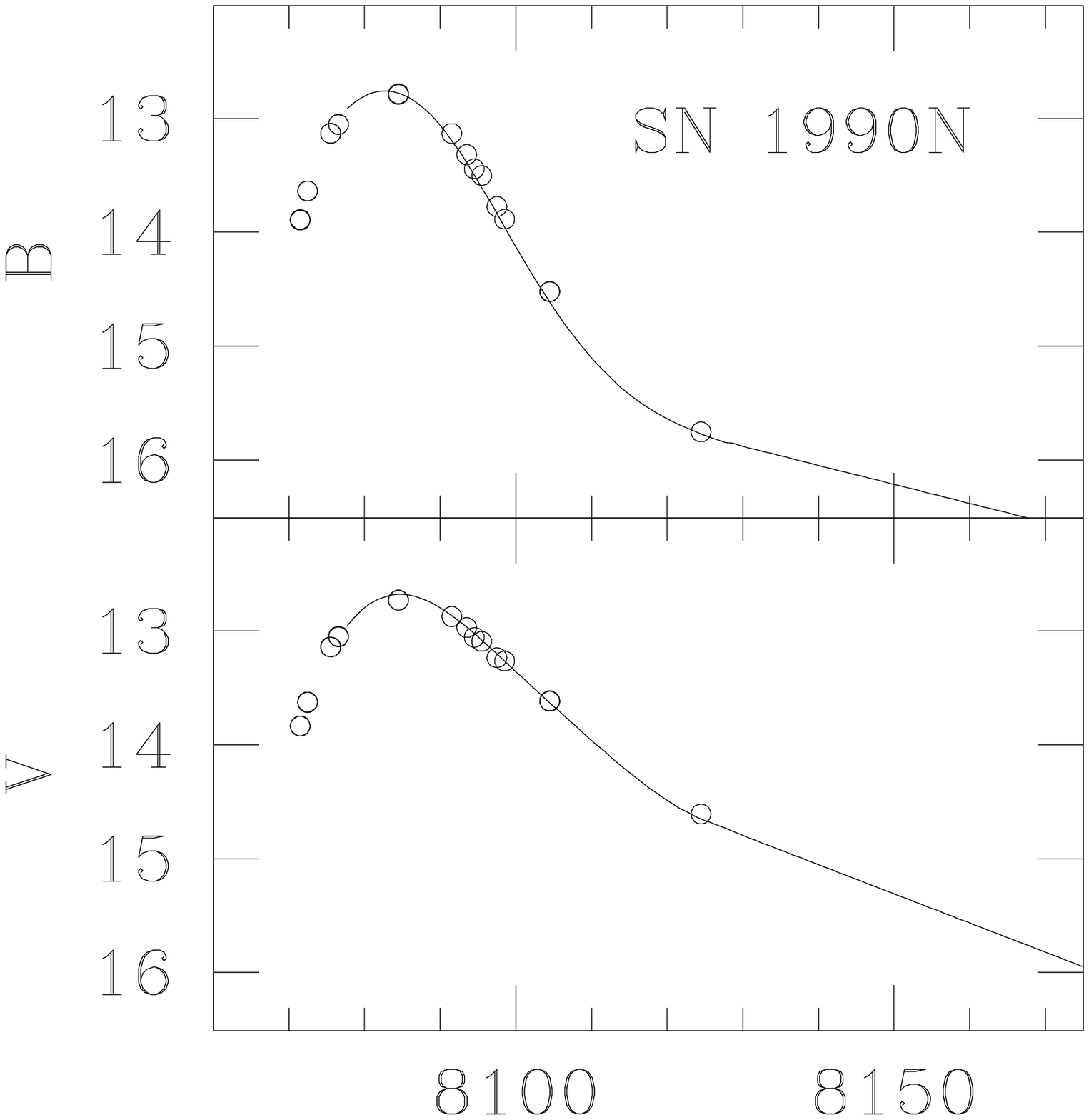]{$BV$ light curves of SN 1990N compared with the SN
Ia template curves determined by \protect{\cite{Lei88}. The abscissa
is plotted in units of days as JD-2440000. }  \label{f7}}
  
\figcaption[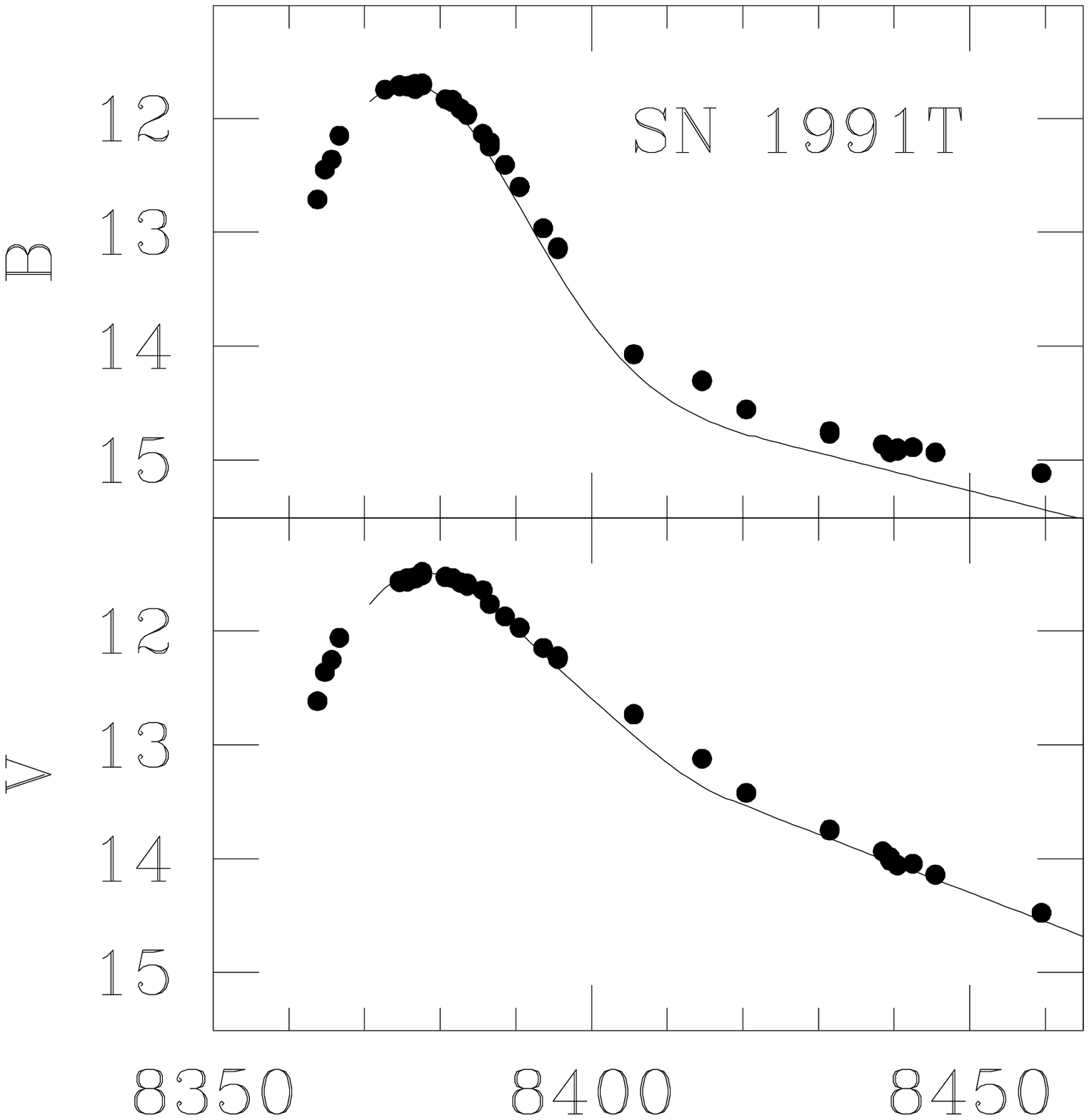]{$BV$ light curves of SN 1991T compared with the SN
Ia template curves determined by \protect{\cite{Lei88} The abscissa is
plotted in units of days as JD-2440000. The data appear to fall from
maximum more slowly than the template curves.}.  \label{f8}}

\figcaption[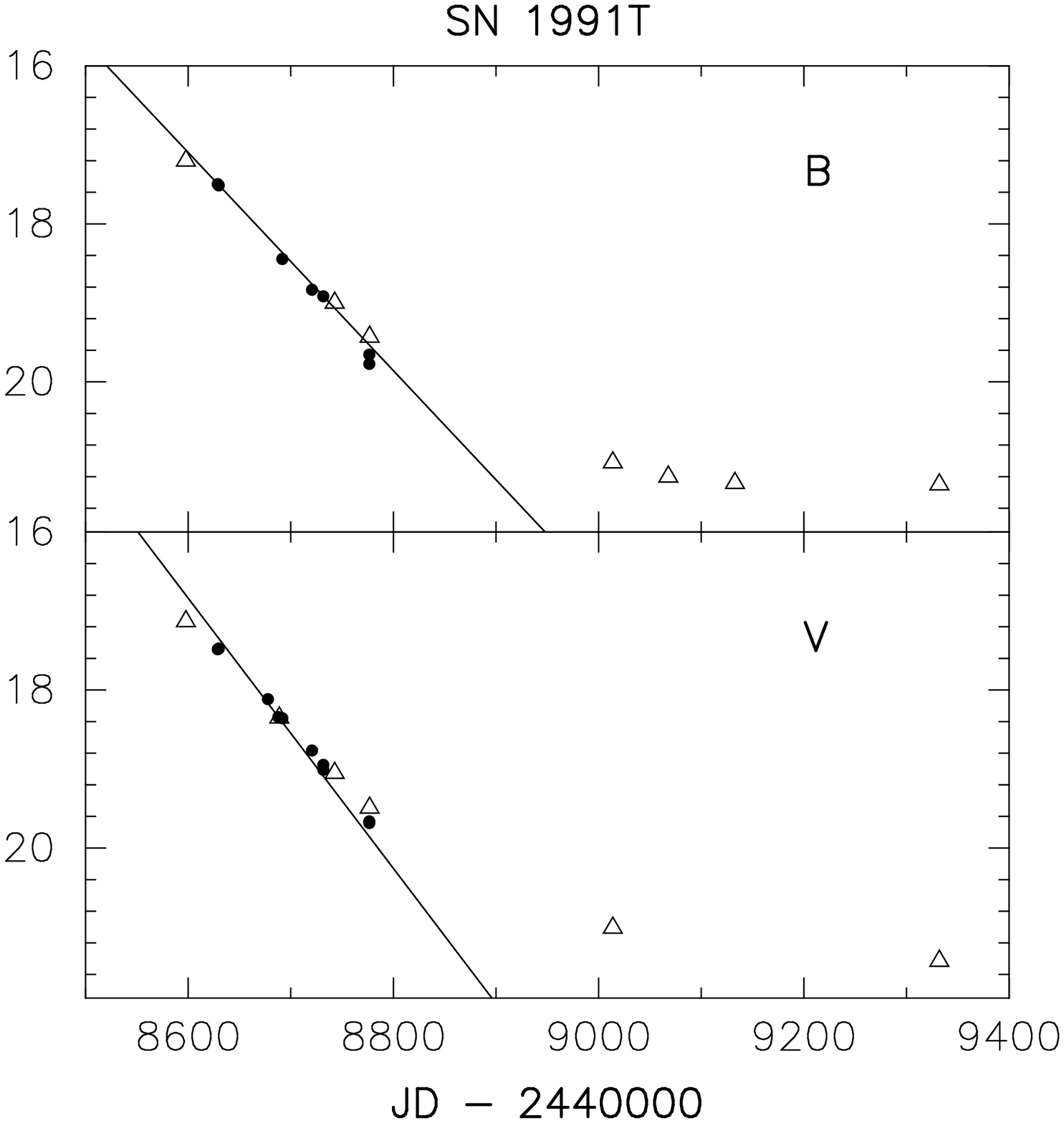]{Late-time $BV$ light curves for SN
1991T. Photometry from CTIO (filled circles) and from KPNO (triangle)
are shown. An estimated decline rate for type Ia supernovae based on
the Pskovskii $\gamma$ parameter is also plotted. \label{f12}}

\figcaption[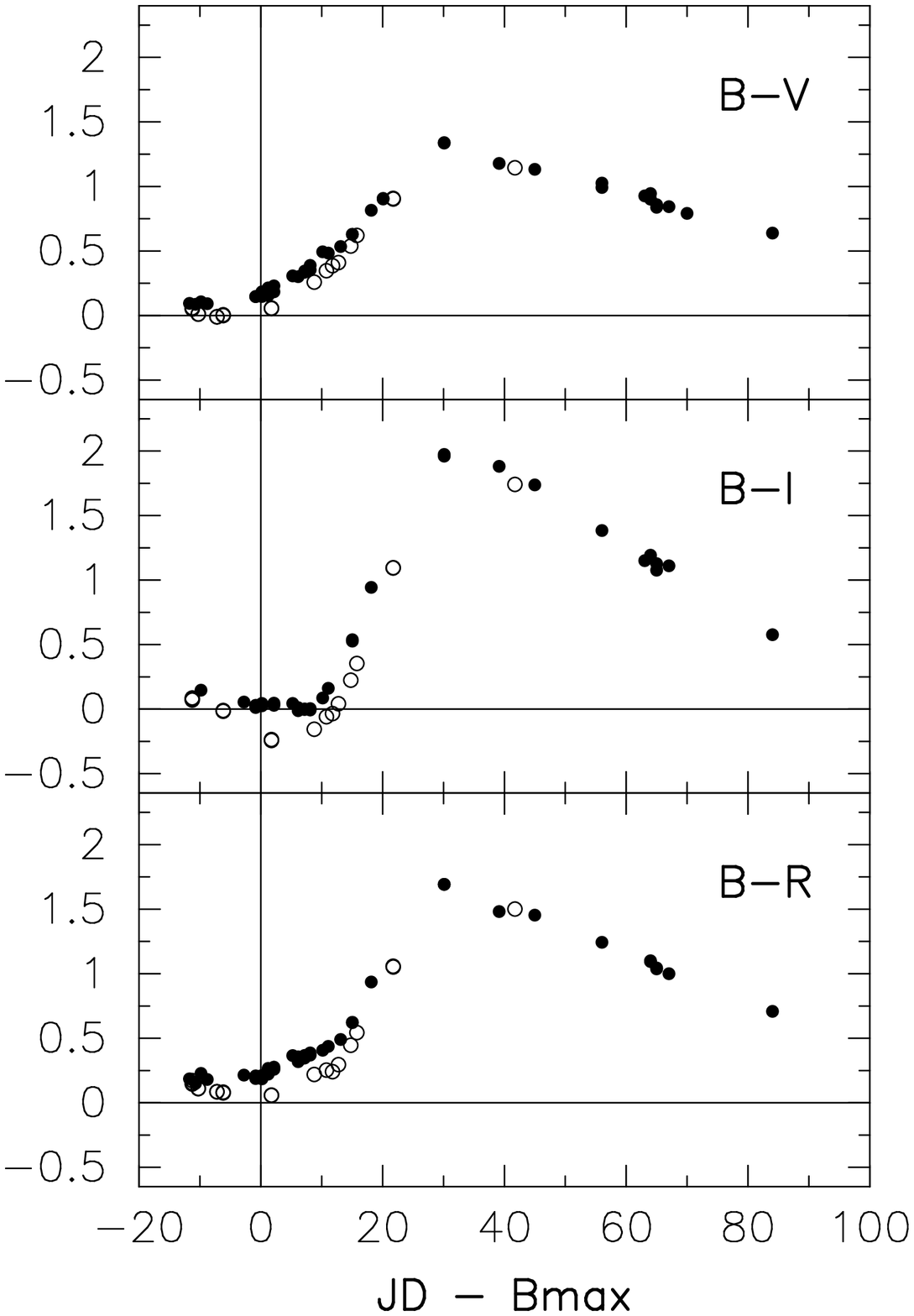]{Observed $B-V$, $B-I$ and $B-R$ color
evolution for SN 1990N (open circles) and SN 1991T (filled circles) as
a function of time since $B_{max}$.   \label{f9}}

\figcaption[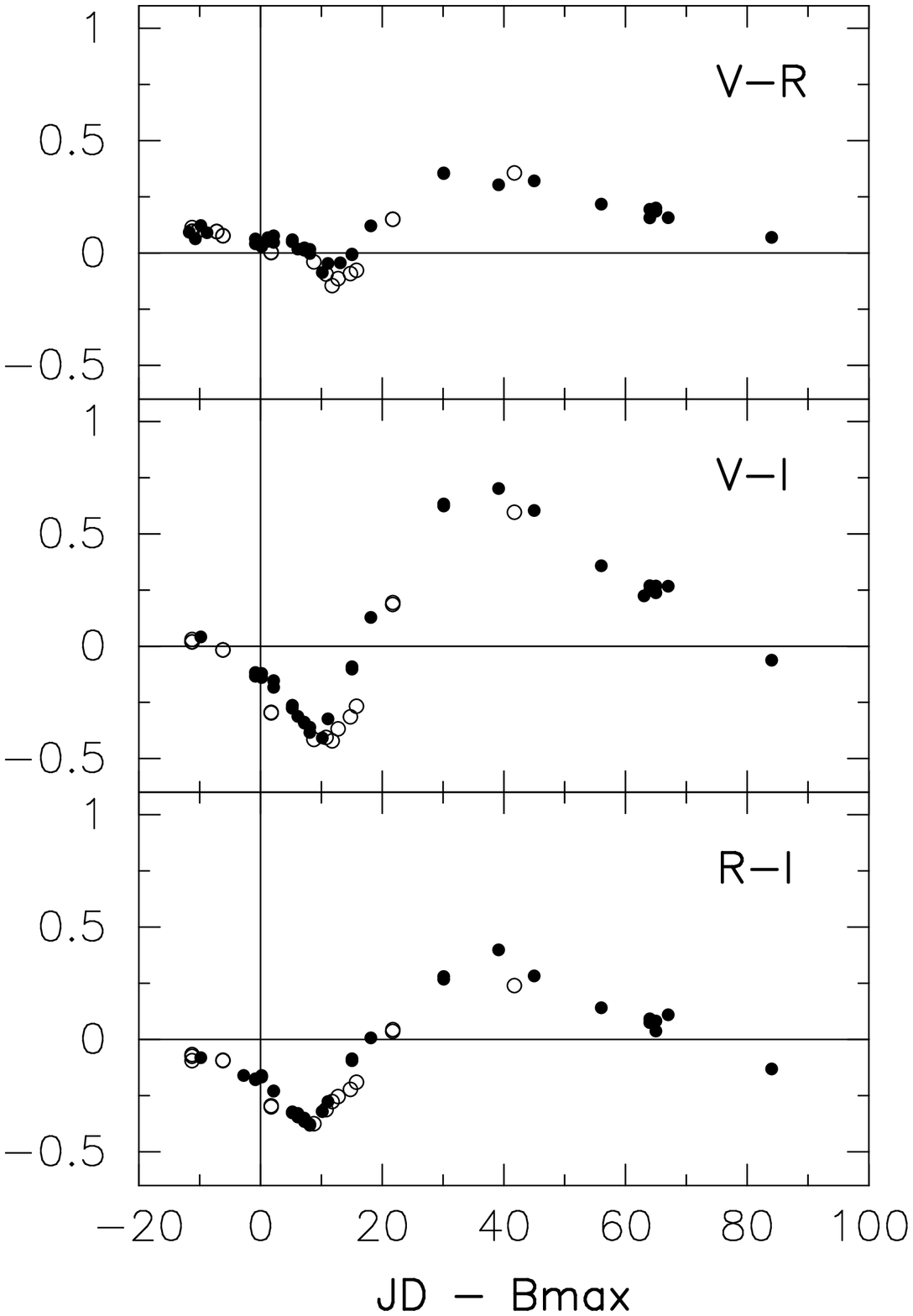]{Observed $V-R$, $V-I$ and $R-I$ color
evolution of SN 1990N (open circles) and SN 1991T (filled circles) as
a function of time since $B_{max}$.  \label{f10}}

\figcaption[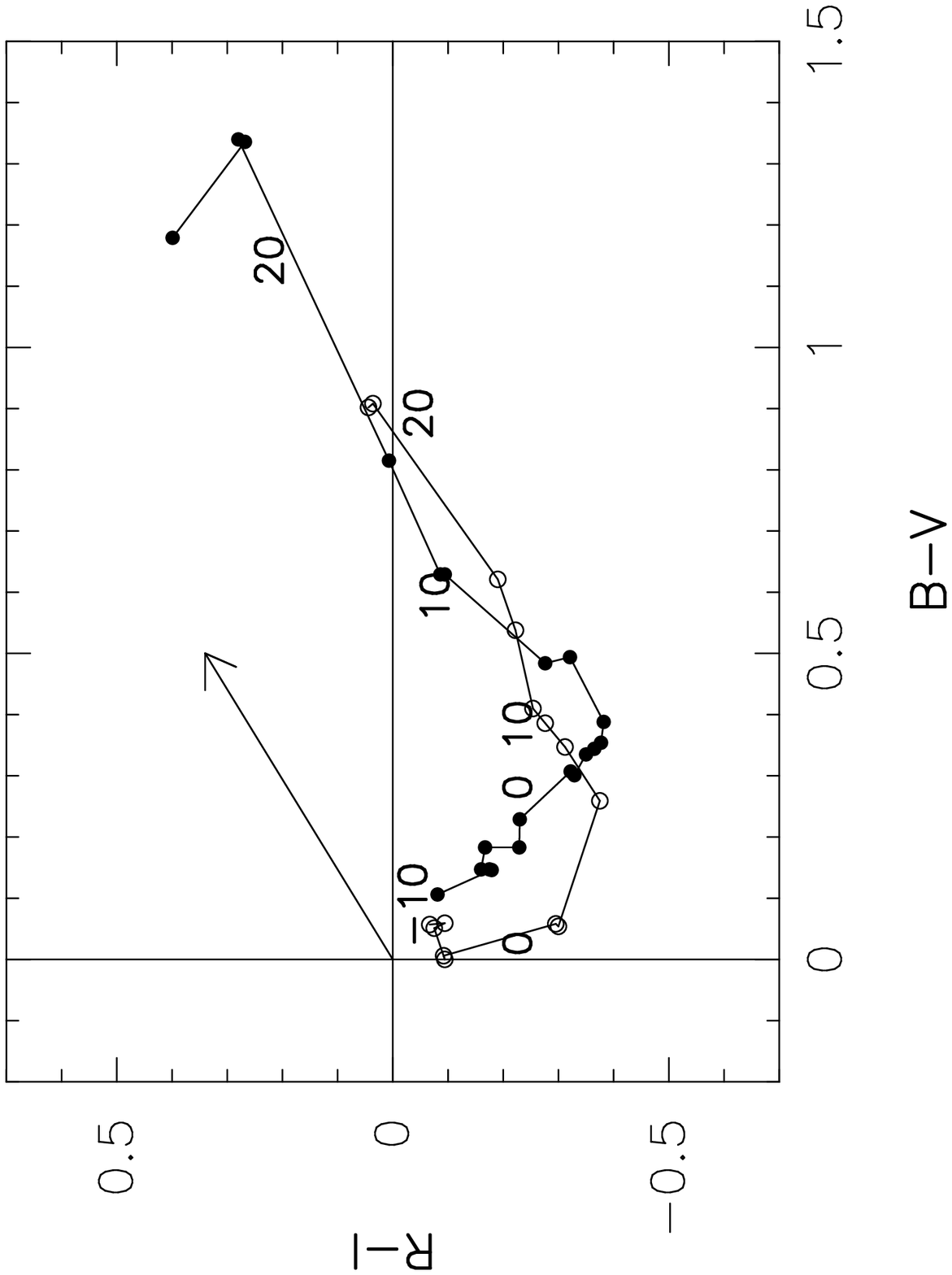]{Observed $B-V$ versus $R-I$ color-color curves
for SN 1990N (circle) and SN 1991T (filled circles). The numbers
indicate the time elapsed, in days, since $B$ maximum. \label{f11}}

\newpage

\begin{table}
\dummytable\label{t1}
\end{table}
{\sc TABLE} \ref{t1}. Observation Log for SN 1990N.

\begin{table}
\dummytable\label{t2}
\end{table}
{\sc TABLE} \ref{t2}. Observation Log for SN 1991T.

\begin{table}
\dummytable\label{t3}
\end{table}
{\sc TABLE} \ref{t3}. Photometric sequence for SN 1990N.

\begin{table}
\dummytable\label{t4}
\end{table}
{\sc TABLE} \ref{t4}. Photometric sequence for SN 1991T.

\begin{table}
\dummytable\label{t5}
\end{table}
{\sc TABLE} \ref{t5}. SN 1990N photometry.

\begin{table}
\dummytable\label{t6}
\end{table}
{\sc TABLE} \ref{t6}. SN 1991T photometry.

\begin{table}
\dummytable\label{t7}
\end{table}
{\sc TABLE} \ref{t7}. Peak photometric magnitudes.

\begin{table}
\dummytable\label{t8}
\end{table}
{\sc TABLE} \ref{t8}. Pskovskii and Phillips parameters.

\clearpage

%
  
\begin{figure}
\plotone{plira.fig3.ps} 
{\center Lira {\it et al.} Figure~\ref{f3}}
\end{figure}
  
\begin{figure}
\plotone{plira.fig4.ps} 
{\center Lira {\it et al.} Figure~\ref{f4}}
\end{figure}

\begin{figure}
\plotone{plira.fig5.ps}
{\center Lira {\it et al.} Figure~\ref{f5}}
\end{figure}

\begin{figure}
\plotone{plira.fig6.ps}
{\center Lira {\it et al.} Figure~\ref{f6}}
\end{figure}

\begin{figure}
\plotone{plira.fig7.ps}
{\center Lira {\it et al.} Figure~\ref{f7}}
\end{figure}
  
\begin{figure}
\plotone{plira.fig8.ps}
{\center Lira {\it et al.} Figure~\ref{f8}}
\end{figure}

\begin{figure}
\plotone{plira.fig9.ps}
{\center Lira {\it et al.} Figure~\ref{f12}}
\end{figure}

\begin{figure}
\plotone{plira.fig10.ps}
{\center Lira {\it et al.} Figure~\ref{f9}}
\end{figure}

\begin{figure}
\plotone{plira.fig11.ps}
{\center Lira {\it et al.} Figure~\ref{f10}}
\end{figure}

\begin{figure}
\plotone{plira.fig12.ps}
{\center Lira {\it et al.} Figure~\ref{f11}}
\end{figure}

\clearpage


\begin{thebibliography}{}


\bibitem[Arnett (1979)]{Arn79} Arnett, W.D. 1979, \apjl, 230, L37

\bibitem[Barbon {\it et al.} 1973]{Bar_etal73} Barbon, R., Ciatti, F., \&
Rosino, L. 1973, \aap, 25, 248

\bibitem[Bertola (1964)]{Ber64} Bertola, F. 1964, Ann.~d'Astroph., 27, 319

\bibitem[Branch (1981)]{Bra81} Branch, D. 1981, \apj, 248, 1076

\bibitem[Branch (1987)]{Bra87} Branch, D. 1987, \apjl, 316, L81

\bibitem[Branch {\it et al.} (1993)]{Bra_etal93} Branch, D., Fisher, A., \&
 Nugent, P. 1993, \aj, 106, 2383

\bibitem[Burstein \& Heiles (1994)]{BurHei94} Burstein, D., \&
 Heiles, C., 1994, \apjs, 54, 33

\bibitem[Cousins (1976)]{Cou76} Cousins, A. W. 1976, \mnras, 81, 25

\bibitem[Filippenko {\it et al.} (1992)]{Fil_etal92} Filippenko,
 A. V., Richmond, M. W., Matheson, T., Shields, J. C., Burbidge,
 E. M., Cohen, R. D., Dickinson, M., Malkan, M. A., Nelson, B., Pietz,
 J., Schlegel, D., Schmeer, P., Spinrad, H., Steidel, C. C., Tran,
 H. D., \& Wren, W.  1992, \apjl, 384, L15

\bibitem[Filippenko (1997)]{Fil97} Filippenko, A. V.,
 1997, \araa, in press.

\bibitem[Filippenko \& Leonard (1995)]{FilLeo95}Filippenko, A.V., 
 \& Leonard, D.C. 1995, \iaucirc 6237

\bibitem[Fisher {\it et al.} (1997)]{Fis_etal97} Fisher, A., Branch,
 D., Nugent, P, \& Baron, E. 1997, \apj, 481, 89

\bibitem[Ford {\it et al.} (1993)]{For_etal93} Ford, C. H., Herbst,
 W., Richmond, M. W., Baker, M. L., Filippenko, A. V., Treffers,
 R. R., Paik, Y., \& Benson, P. J. 1993, \aj, 106, 1101

\bibitem[Garvavich {\it et al.} (1996)]{Gar_etal96} Garnavich, P.M.,
 Riess, A.G., Kirshner, R.P., Challis, P., \& Wagner, R.M.  1996,
 \baas, 28, 1331

\bibitem[Grothues \& Gochermann (1992)]{GroGoc92} Grothues, H.-G., \&
Gochermann, J. 1992, ESO Messenger, 68, 43

\bibitem[Graham (1982)]{Gra82} Graham, J. A. 1982, \pasp, 94, 244

\bibitem[Graham {\it et al.} (1997)]{Gra_etal97} Graham, J. A.,
 Phelps, R. L., Freedman, W. L., Saha, A., Ferrarese, L., Stetson,
 P. B., Madore, B. F., Silbermann, N. A., Sakai, S., Kennicutt, R. C.,
 Harding, P., Bresolin, F., Turner, A., Mould, J. R., Rawson, D. M.,
 Ford, H. C., Hoessel, J. G., Han, M., Huchra, J. P., Macri, L. M.,
 Hughes, S. M., Illingworth, G. D., \& Kelson, D. D.  1997, \apj, 477,
 535

\bibitem[Hamuy {\it et al.} (1991)]{Ham_etal91} Hamuy, M., Phillips,
 M. M., Maza, J., Wischnjewsky, M., Uomoto, A., Landolt, A. U., \&
 Khatwani, R.  1991, \aj, 102, 208

\bibitem[Hamuy {\it et al.} (1994)]{Ham_etal94} Hamuy, M., Phillips,
 M. M., Maza, J., Suntzeff, N. B., Della Valle, M., Danziger, J.,
 Antezana, R., Wischnjwesky, M., Aviles, R., Schommer, R. A., Kim,
 Y.-C., Wells, L. A., Ruiz, M. T., Prosser, C. F., Krzeminski, W.,
 Baylin, C. D., Hartigan, P., \& Hughes, J.  1994, \aj, 108, 2226

\bibitem[Hamuy {\it et al.} (1995)]{Ham_etal95} Hamuy, M., Phillips, M. M.,
 Maza, J., Suntzeff, N. B., Schommer, R. A., \& Avil\'es, R. 1995, \aj,
 109, 1

\bibitem[Hamuy {\it et al.} (1996a)]{Ham_etal96a} Hamuy, M., Phillips,
 M. M., Schommer, R. A., Suntzeff, N.B., Maza, J., \& Avil\'es, R.
 1996a, \aj, 112, 2391 

\bibitem[Hamuy {\it et al.} (1996b)]{Ham_etal96b} Hamuy, M.,
 Phillips, M. M., Suntzeff, N. B., Schommer, R. A., Maza, J., \&
 Avil\'es, R.  1996b, \aj, 112, 2398

\bibitem[Hamuy {\it et al.} (1996c)]{Ham_etal96c} Hamuy, M., Phillips,
 M. M., Suntzeff, N. B., Schommer, R. A., Maza, J., Smith, R. C., Lira
 P., \& Avil\'es, R.  1996c, \aj, 112, 2438 

\bibitem[Harris {\it et al.} (1981)]{Har_etal81} Harris, W. E.,
 Fitzgerald, H. P., \& Reed, B. C. 1981, \pasp, 93 507

\bibitem[Harkness \& Wheeler (1990)]{HarWhe90} Harness, R.P., and
 Wheeler, J.C. 1990, in Supernovae, ed. Petschek, A.G., New York:
 Springer-Verlag, p. 1.

\bibitem[Jacoby \& Pierce (1996)]{JacPie96} Jacoby, G. H., \& Pierce,
 M. J. 1996, \aj, 112, 723

\bibitem[Jeffery {\it et al.} (1992)]{Jef_etal92} Jeffery, D. J.,
 Leibundgut, B., Kirshner, R. P., Benetti, S., Branch, D., \&
 Sonneborn, G. 1992, \apj, 397, 304

\bibitem[Johnson (1963)]{Joh63} Johnson, H. L., 1963, in Basic
 Astronomical Data, edited by K. A. Strand, (Univ. Chicago Press,
 Chicago), p. 204

\bibitem[Kirshner \& Leibundgut (1990)]{KirLei90} Kirshner, R., and
 Leibundgut, B. 1990, \iaucirc, No. 5039

\bibitem[Kirshner {\it et al.} (1991)]{Kir_etal91} Kirshner, R. P.,
 Leibundgut, B, Foltz, C. B. , \& Pier, J. 1990, \apj, 118, 502

\bibitem[Kirshner (1991)]{Kir91} Kirshner R. P., 1991, \iaucirc No 5239

\bibitem[Kron (1953)]{Kro53} Kron, R. G. 1953, \apj, 118, 502

\bibitem[La Franca \& Goldschmidt (1991)]{LaFGol91} La Franca, F., \& 
 Goldschmidt, C. 1991, \iaucirc No. 5239

\bibitem[Landolt (1972)]{Lan72} Landolt, A. U. 1972, \aj, 78, 959

\bibitem[Landolt (1992)]{Lan92} Landolt, A. U. 1992, \aj, 104, 340

\bibitem[Leibundgut (1988)]{Lei88} Leibundgut, B. 1988, Ph.D. Thesis, 
 Universitat Basel

\bibitem[Leibundgut {\it et al.} (1991)]{Lei_etal91} Leibundgut, B.,
 Kirshner, R. P., Filippenko, A. V, Shields, J. C., Foltz, C. B.,
 Phillips, M. M., \& Sonneborn, G. 1991, \apjl, 371, 23

\bibitem[Lira (1996)]{Lir96} Lira P. 1996, Mc.S. thesis, Universidad
 de Chile

\bibitem[Madore \& Freedman (1991)]{MadFre91} Madore, B.F., and
 Freedman, W.L. 1991, \pasp, 103, 933

\bibitem[Maza {\it et al.} (1994)]{Maz_etal94} Maza, J., Hamuy, M.,
 Phillips, M. M., Suntzeff, N. B., \& Avil\'es R. 1994, \apjl, 424,
 L107

\bibitem[Mazzali {\it et al.} (1995)]{Maz_etal95} Mazzali, P. A.,
 Danziger, I. J., \& Turatto, M. 1995, \aa, 297, 509

\bibitem[Maury (1990)]{Mau90} Maury, A. 1990, \iaucirc, No. 5039

\bibitem[Meikle {\it et al.} (1996)]{Mei_etal96} Meikle, W. P. S.,
 Cumming, R. J., Geballe, T. R., Lewis, J. R., Walton, N. A.,
 Balcells, M., Cimatti, A., Croom, S. M., Dhillon, V. S., Economou,
 F., Jenkins, C. R., Knapen, J. H., Meadows, V. S., Morris, P. W.,
 Perez-Fournon, I., Shanks, T., Smith, L. J., Tanvir, N. R., Veilleux,
 S., Vilchez, J., Wall, J. V., \& Lucey, J. R. 1996, \mnras, 281, 263


\bibitem[Meyer \& Roth (1991)]{MeyRot91} Meyer, D.M., \& Roth, K.C.
 1991, \apjl, 383, L41

\bibitem[Nomoto {\it et al.} (1994)]{Nom_etal84} Nomoto, K.,
Thielemann, F.-K., \& Yokoi, K. 1984, \apj, 286, 644

\bibitem[Nugent {\it et al.} (1995)]{Nug_etal95} Nugent, P., Phillips,
 M. M., Baron, E., Branch, D., \& Hauschildt, P. 1995, \apjl, 455,
 L147

\bibitem[Olsen (1983)]{Ols83} Olsen, E.H. 1973, \aaps, 54, 55

\bibitem[Pierce \& Jacoby (1995)]{PieJac95} Pierce, M. J., \& Jacoby,
 G. H. 1995, \aj, 110, 2885

\bibitem[Phillips {\it et al.} (1987)]{Phi_etal87} Phillips, M. M.,
 Phillips, A. C., Heathcote, S. R., Blanco, V. M., Geisler, D.,
 Hamilton, D., Suntzeff, N. B., Jablonski, F. J., Steiner, J. E.,
 Cowley, A. P. {\it et al.} 1987, \pasp, 99, 592


\bibitem[Phillips (1993)]{Phi93} Phillips, M. M. 1993, \apjl, 413,
 L105

\bibitem[Phillips {\it et al.} (1992)]{Phi_etal92} Phillips, M. M.,
 Suntzeff, N. B., Hamuy, M., Leinbundgut, B., Kirshner, R. P., \&
 Foltz, C. B. 1992, \aj, 103, 1632

\bibitem[Phillips \& Hamuy (1991)]{PhiHam91} Phillips, M. M., \&
 Hamuy, M.  1991, \iaucirc No. 5251


\bibitem[Pollas (1990)]{Pol90} Pollas, C. 1990, \iaucirc No. 5040

\bibitem[Pskovskii (1977)]{Psk77} Pskovskii, Y. P. 1977, Soviet Astr,
 21, 675

\bibitem[Pskovskii (1984)]{Psk84} Pskovskii, Y. P. 1984, Soviet Astr.,
 28, 658

\bibitem[Qiao {\it et al.} (1997a)]{Qia_etal97a} Qiao, Q.-Y., Wu, H.,
 Wei, J.-Y., \& Li, W.-D. 1997a, \iaucirc 6623  

\bibitem[Qiao {\it et al.} (1997b)]{Qia_etal97b} Qiao, Q.-Y., Wu, H., 
 Wei, J.-Y., \& Li, W.-D. 1997b, \iaucirc, 6642 

\bibitem[Riess {\it et al.} (1995)]{Rie_etal95} Riess, A. G., Press,
 W. H., \& Kirshner, R. P. 1995, \apjl, 438, L17

\bibitem[Riess {\it et al.} (1996)]{Rie_etal96} Riess, A. G., Press,
 W. H., \& Kirshner, R. P. 1996, \apj, 473, 88

\bibitem[Ruiz-Lapuente et al (1992)]{Rui_etal92} Ruiz-Lapuente, P.,
 Cappellaro, E., Turatto, M., Gouiffes, C., Danziger, I. J., Della
 Valle, M., \& Lucy, L. B. 1992, \apjl, 387, L33

\bibitem[Rust (1974)]{Rus74} Rust, B.W., ``The Use of Supernovae Light
 Curves for Testing the Expansion Hypothesis and Other Cosmological
 Relations,'' Ph.D. thesis, Univ. Illinois.

\bibitem[Saha {\it et al.} (1996)]{Sah_etal96} Saha, A., Sandage, A.,
 Labhardt, L., Tammann, G. A., Macchetto, F. D., \& Panagia, N. 1996,
 \apjs, 107, 693

\bibitem[Sandage \& Tammann (1993)]{SanTam93} Sandage, A., \& Tammann,
 G. A. 1993, \apj, 415, 1

\bibitem[Sandage {\it et al.} (1996)]{San_etal96} Sandage, A., Saha,
 A., Tammann, G. A., Labhardt, L., Panagia, N., \& Maccheto,
 F. D. 1996, \apjl, 460, L15

\bibitem[Savage \& Mathis (1979)]{SavMat79} Savage, B. D., \& Mathis,
 J. S. 1979, \araa, 17, 73

\bibitem[Schaefer (1996)]{Sch96} Schaefer, B. 1996, \aj, 111, 1668

\bibitem[Schmidt {\it et al.} (1994)]{Sch_etal94} Schmidt, B. P.,
 Kirshner, R. P., Leibundgut, B., Wells, L. A., Porter, A. C.,
 Ruiz-Lapuente, P., Challis, P., \& Filippenko, A. V. 1994, \apjl,
 434, L19

\bibitem[Serkowski (1970)]{Ser70} Serkowski, K. 1970, \pasp, 82, 908

\bibitem[Shigeyama {\it et al.} (1992)]{Shi_etal92} Shigeyama, T.,
 Nomoto, K., Yamaoka, H., \& Thielemann, F. 1992, \apjl, 386, L13

\bibitem[Silbermann {\it et al.} (1996)]{Sil_etal96} Silbermann, N.A.,
Harding, P., Madore, B.F., Kennicutt, R.C., Freedman, W.L., Mould,
J.R., Stetson, P.B., Saha, A., Bresolin, F., Turner, A., Ferrarese,
L., Ford, H., Gibson, B., Rawson, D., Graham, J., Han, M., Hoessel,
J.G., Hill, R.J., Huchra, J., Macri, J., Hughes, S.M.G., Illingworth,
G.D., Kelson, D., Phelps, R., \& Sakai, S.  1996, In `` The STScI May
Symp.; The Extragalactic Distance Scale'' STScI preprint of poster
papers, p. 67

\bibitem[Spyromilio {\it et al.} (1992)]{Spy_etal92} Spyromilio, J.,
 Meikle, W. P., Allen, D. A., \& Graham, J. R. 1992, \mnras, 258, 53

\bibitem[Stetson (1987)]{Ste87} Stetson, P. B. 1987, \pasp, 99, 191

\bibitem[Suntzeff (1995)]{Sun95} Suntzeff, N. B. 1995, in Supernovae
 and Supernova Remnants, IAU Colloquium 145, edited by R. McCray
 (Cambridge University Press, Cambridge), p. 41

\bibitem[Tammann \& Sandage (1995)]{TamSan95} Tammann, G. A., \& Sandage,
 A., 1995, \apj, 452, 16

\bibitem[Tanvir {\it et al.} (1995)]{Tan_etal95} Tanvir, N. R., Shanks,
 T., Ferguson, H. C., Robinson, D. T. R. 1995, \nat, 377, 27

\bibitem[Waagen {\it et al.} (1991)]{Waa_etal91} Waagen, E. {\it et
 al.} 1991, \iaucirc No. 5239

\bibitem[Walker {\it et al.} (1970)]{Wal_etal70} Walker, G.A.H.,
 Andrews, D., Hill, G., Morris, S.C., Smyth, W. and White, J., 1970,
 D.A.O.  Publications XIII, Pt. 17, 415

\bibitem[Wells {\it et al.} 1994]{Wel_etal94} Wells, L. A., Phillips,
 M. M., Suntzeff, B., Heathcote, S. R., Hamuy, M., Navarrete, M.,
 Fernandez, M., Weller, W. G., Schommer, R. A., Kirshner, R. P.,
 Leibundgut, B., Willner, S. P., Peletier, S. P., Schlegel, E. M.,
 Wheeler, J. C., Harkness, R. P., Bell, D. J., Matthews, J. M.,
 Filippenko, A. V., Shields, J. C., Richmond, W., Jewitt, D., Luu, J.,
 Tran, H. D., Appleton, P. N., Robson, E. I., Tyson, J. A.,
 Guhathakurta, P., Eder, J. A., Bond, H. E., Potter, M., Veilleux, S.,
 Porter, A. C., Humphreys, R. M., Janes, K. A., Williams, T. B.,
 Costa, E., Ruiz, M. T., Lee, J. T., Lutz, J. H., Rich, R. M.,
 Winkler, P. F., \& Tyson, N. D.  1994, \aj, 108, 2233

\bibitem[Wheeler \& Smith (1991)]{WheSmi91} Wheeler, J.C., \& Smith,
V.V.  1991, \iaucirc, 5256

\bibitem[Woosley {\it et al.} 1989 ]{Woo_etal89} Woosley, S.E., Pinto,
P.A., Hartmann, D. 1989, \apj, 346, 395

\bibitem[Yamaoka {\it et al.} (1992)]{Yam_etal92} Yamaoka, H., Nomoto,
 K., Shigeyama, T., \& Thielemann, F. 1992, \apjl, 393, L55


\end{thebibliography}
\end{document}